\documentclass[iop,apj,appendixfloats]{emulateapj}
\usepackage{apjfonts}
\usepackage{graphicx}
\usepackage{amsmath}
\usepackage{amssymb}
\gdef\xx[#1]{\textcolor{red}{#1}}

\gdef\kms{km\,s$^{-1}$}
\gdef\msun{M$_{\odot}$}

\lefthead{van Dokkum et al.}
\righthead{}
\slugcomment{Accepted for publication in ApJ Letters}
\begin{document}

\title{Extensive globular cluster systems associated with ultra
diffuse galaxies in the Coma cluster}

\author{Pieter van Dokkum\altaffilmark{1}, Roberto Abraham\altaffilmark{2},
Aaron J.\ Romanowsky\altaffilmark{3,4},
Jean Brodie\altaffilmark{3}, Charlie
Conroy\altaffilmark{5}, Shany
Danieli\altaffilmark{1}, Deborah Lokhorst\altaffilmark{2},
Allison Merritt\altaffilmark{1},
Lamiya Mowla\altaffilmark{1}, 
Jielai
Zhang\altaffilmark{2}
\vspace{8pt}}

\altaffiltext{1}
{Astronomy Department, Yale University, New Haven, CT 06511, USA}
\altaffiltext{2}
{Department of Astronomy \& Astrophysics, University of Toronto,
   50 St.\ George Street, Toronto, ON M5S 3H4, Canada}
\altaffiltext{3}
{University of California Observatories, 1156 High Street, Santa
Cruz, CA 95064, USA}
\altaffiltext{4}
{Department of Physics and Astronomy, San Jos\'e State University,
San Jose, CA 95192, USA}
\altaffiltext{5}
{Harvard-Smithsonian Center for Astrophysics, 60 Garden Street,
Cambridge, MA, USA}

\begin{abstract}
We present {\em Hubble Space Telescope}
imaging of two 
ultra diffuse galaxies (UDGs) with
measured stellar velocity dispersions
in the Coma cluster.
The galaxies, Dragonfly\,44 and DFX1, have
effective radii of $4.7$\,kpc and $3.5$\,kpc and
%central surface brightness $\mu_{0,V}=24.1$\,mag\,arcsec$^{-2}$
%and $\mu_{0,V}=24.0$\,mag\,arcsec$^{-2}$, and velocity dispersions
velocity dispersions of
$47^{+8}_{-6}$\,\kms\ and
$30^{+7}_{-7}$\,\kms, respectively.
Both galaxies are associated with a striking number of
compact objects, tentatively identified as globular clusters:
$N_{\rm gc}=74\pm18$ for Dragonfly\,44 and $N_{\rm gc}=62\pm 17$ for DFX1.
%They are blue, with $\langle V_{606}-I_{814}\rangle = XXX$.
The number of globular clusters
is much
higher than expected from the luminosities of the galaxies but is
consistent with expectations from
the empirical relation between dynamical mass and globular
cluster count defined by other galaxies.
Combining our data 
with previous {\em HST} observations of Coma UDGs
we find that UDGs have a factor of $6.9^{+1.0}_{-2.4}$
more globular clusters than other galaxies of the same
luminosity,
in contrast to a recent study of a similar sample by
Amorisco et al.\ (2017), but consistent with
earlier results for individual
galaxies.
The Harris et al.\ (2017) relation between
globular cluster count and dark matter halo mass implies a median
halo mass of $M_{\rm halo}\sim 1.5 \times 10^{11}$\,\msun\ for
the sixteen Coma UDGs that have been observed with {\em HST} so far,
with the largest and brightest having $M_{\rm halo}\sim
5\times 10^{11}$\,\msun.

\end{abstract}


\section{Introduction}

The discovery of large, extremely faint, spheroidal objects in galaxy
clusters dates at least to {Impey}, {Bothun}, \& {Malin} (1988),
who noticed several
such objects in photographic studies of the Virgo cluster. Over the
following three decades several more were found (e.g., {Dalcanton} {et~al.} 1997),
but it was only recognized recently how common they are.
Using the Dragonfly Telephoto Array ({Abraham} \& {van Dokkum} 2014),
47 galaxies with half-light radii $R_{\rm e}\gtrsim 1.5$\,kpc
and central surface brightness $\mu_{g,0}\gtrsim 24$\,mag\,arcsec$^{-2}$
were found in the Coma cluster ({van Dokkum} {et~al.} 2015a).
The galaxies appear smooth and spheroidal, and have a much lower
{Sersic} (1968) index than elliptical galaxies ($n\sim 1$ versus
$n\sim 4$ for ellipticals).
These remarkable objects were dubbed ``ultra diffuse galaxies'',
or UDGs. The number of known UDGs quickly expanded in the past two
years, with many more
examples found in Coma ({Koda} {et~al.} 2015), Virgo ({Mihos} {et~al.} 2015),
other clusters ({van der Burg}, {Muzzin}, \&  {Hoekstra} 2016), and in low density environments
({Mart{\'{\i}}nez-Delgado} {et~al.} 2016; {Merritt} {et~al.} 2016; {Rom{\'a}n} \& {Trujillo} 2017).

It is still unknown how UDGs fit in the general framework of galaxy
formation and evolution. One possibility is that most UDGs are closely
related to smaller galaxies of the same luminosity:
they may have originated as small
galaxies that were puffed up by tidal interactions
(see, e.g., {Collins} {et~al.} 2013), or represent the high angular momentum
tail of the general population of dwarf galaxies
({Amorisco} \& {Loeb} 2016). Another possibility is that many UDGs are ``failed''
galaxies, with truncated star formation histories. Strong feedback
from supernovae or active nuclei
could produce underluminous galaxies,
perhaps in combination with environmental effects
({Agertz} \& {Kravtsov} 2015; {Yozin} \& {Bekki} 2015; {Di Cintio} {et~al.} 2017). 

Intriguingly, an important clue to the formation of these 
diffuse galaxies
comes from the most compact stellar systems in the universe.
{Beasley} {et~al.} (2016) found that the UDG VCC\,1287 in Virgo has a surprisingly
large number
of globular clusters for its luminosity.
Similar results were subsequently
reported
for the Coma UDGs Dragonfly~17 ({Peng} \& {Lim} 2016) and Dragonfly~44
({van Dokkum} {et~al.} 2016).
%As there is an empirical
%relation between the number of globular clusters and the dark matter
%halo mass of galaxies ({Harris}, {Harris}, \& {Alessi} 2013; {Harris}, {Blakeslee}, \&  {Harris} 2017), these values may
%imply relatively high halo masses of $1-7 \times 10^{11}$\,\msun.
%High halo masses are also implied by the first kinematic measurements
%of UDGs ({Beasley} {et~al.} 2016; {van Dokkum} {et~al.} 2016), and by their
%structural integrity in the harsh environment of rich clusters
%({van Dokkum} {et~al.} 2015a; {Yozin} \& {Bekki} 2015; {van der Burg} {et~al.} 2016).
These early results, together with the first measurements of the kinematics
of UDGs ({Beasley} {et~al.} 2016; {van Dokkum} {et~al.} 2016),
indicated that UDGs are fundamentally different
from other galaxies of the same luminosity.

However, other
studies have cast doubt on this interpretation.
Some large, low surface brightness objects seem to be
tidally-disrupted low mass galaxies
({Collins} {et~al.} 2013; {Merritt} {et~al.} 2016), and there is large variation
in the cold gas fraction among field UDGs ({Papastergis}, {Adams}, \&  {Romanowsky} 2017).
Furthermore,
it has been suggested that massive, globular cluster-rich systems
are the exception, not the rule:
{Amorisco}, {Monachesi}, \&  {White} (2017) report that UDGs have {\em no}
statistically-significant
excess of globular clusters compared to normal dwarf galaxies with
the same stellar mass. Amorisco et al.\ come to this conclusion
from a comparison of
the positions of compact objects in the
{Hammer} {et~al.} (2010) {\em HST}/ACS
Coma Cluster Treasury program (CCTp; Carter et al.\ 2008)
catalog to the positions of
low surface brightness objects in the {Yagi} {et~al.} (2016) catalog.

\begin{figure*}[htbp]
  \begin{center}
  \includegraphics[width=0.89\linewidth]{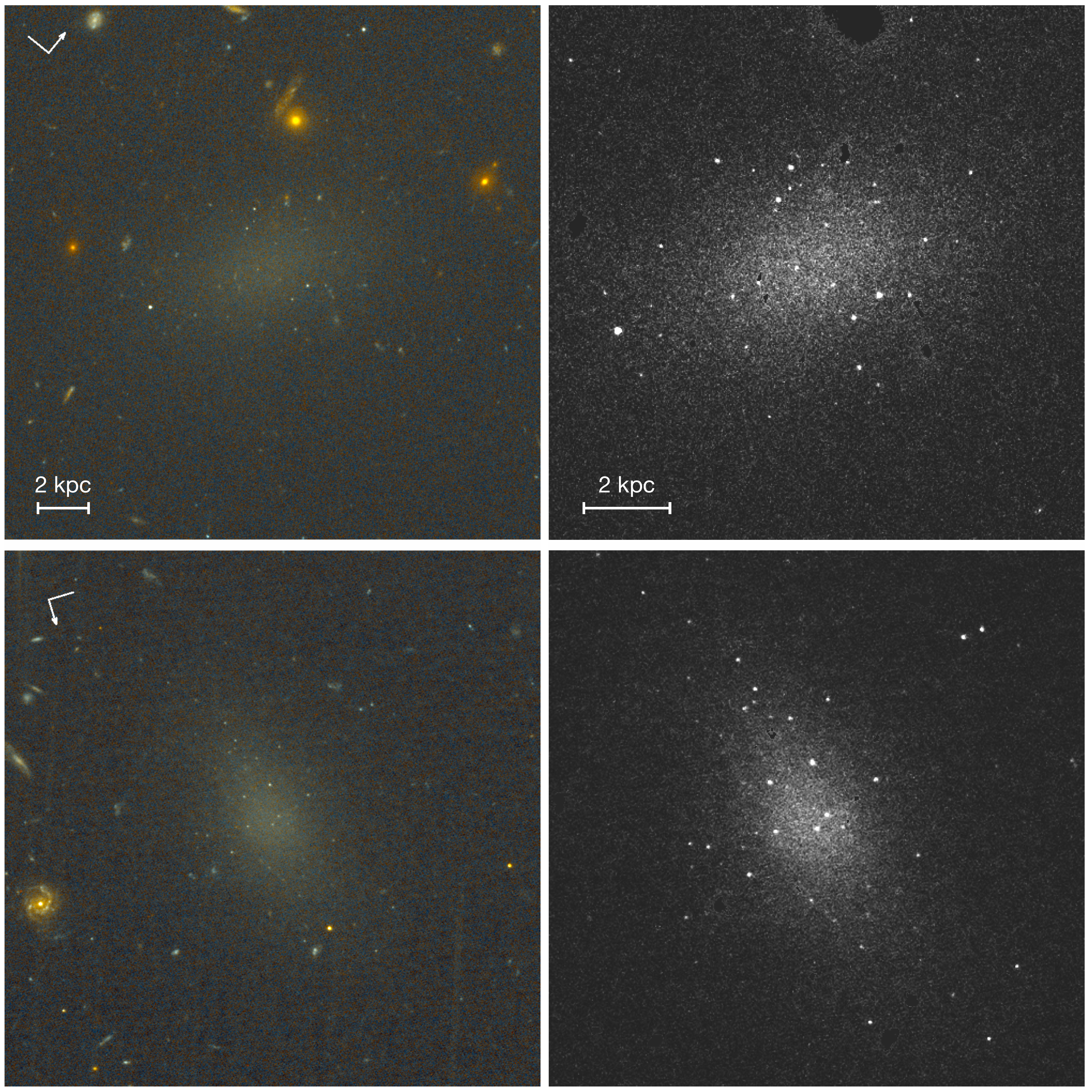}
  \end{center}
\vspace{-0.2cm}
    \caption{
{\em HST} images of 
Dragonfly\,44 (top) and DFX1 (bottom).
The left panels span $42\arcsec \times 42\arcsec$ ($20\,{\rm kpc}\times
20\,{\rm kpc}$) and
were created from the $V_{606}$ and $I_{814}$ images.
The right panels span $25\arcsec \times 25\arcsec$ ($12\,{\rm kpc}\times
12\,{\rm kpc}$) and show the deep $V_{606}$ data at higher contrast,
with spatially-extended objects masked (see text).
Both galaxies are associated with a large number of compact
objects, identified as globular clusters.
}
\label{image.fig}
\end{figure*}

In this {\em Letter} we contribute to this discussion
by measuring the globular cluster
populations in two large
Coma UDGs using {\em HST}. We also
analyze all archival {\em HST}/ACS images
of UDGs in Coma.

\section{Data}
\subsection{Kinematics}

In 2016 we obtained very deep spectroscopy of several
low surface brightness objects in the Coma cluster with the Deep Imaging
Multi-Object Spectrograph (DEIMOS) on the Keck II telescope.
The primary targets were Dragonfly\,44, one of the largest UDGs in Coma,
and a similar-looking galaxy that we dubbed DFX1.
The latter object was visually identified in an archival CFHT/Megacam image;
its J2000 coordinates are $\alpha=13^h01^m15.8^s$,
$\delta=27^{\circ}12'37''$ and it is listed in various previous
catalogs (2175 in Godwin, Metcalfe, \& Peach 1983;
13 in {Yagi} {et~al.} 2016).
%{Godwin} {et~al.} (1983) catalog (\#2175).
It was not in the original Dragonfly UDG
catalog as we removed all objects that
were detected in the Sloan Digital Sky Survey.
We also obtained a spectrum of Dragonfly\,42, a very faint UDG.
%from
%{van Dokkum} {et~al.} (2015a).

The instrumental resolution
($\sigma_{\rm instr}=32$\,\kms) and exposure time
(120,600\,s) were sufficient for measuring
the central
stellar velocity dispersions of both Dragonfly\,44 and DFX1.
For  Dragonfly\,44 we measure
$\sigma=47^{+8}_{-6}$\,\kms,
as described in {van Dokkum} {et~al.} (2016).
Using the same methodology 
we find 
$\sigma=30^{+7}_{-7}$\,\kms\ for DFX1. Its redshift is
$z=0.02741\pm{}0.00002$. For Dragonfly\,42 we could only measure
the redshift: $z=0.02122\pm{}0.00007$.
DFX1 and Dragonfly\,42
contribute to the steadily growing sample of UDGs with confirmed distances
(see {Kadowaki}, {Zaritsky}, \&  {Donnerstein} 2017), and our redshifts confirm that
Dragonfly\,44, DFX1, and Dragonfly\,42 are all members of the
Coma cluster.\footnote{The redshift of Dragonfly\,44 was
not listed in van Dokkum et al.\ (2016), except erroneously
inside Fig.\ 2 of that paper. The correct redshift is
$z=0.02132 \pm 0.00002$.
Dragonfly\,42 and Dragonfly\,44 are likely
bound, as their radial velocities are less than 
$50$\,\kms\ apart.}

\begin{figure*}[htbp]
  \begin{center}
  \includegraphics[width=0.85\linewidth]{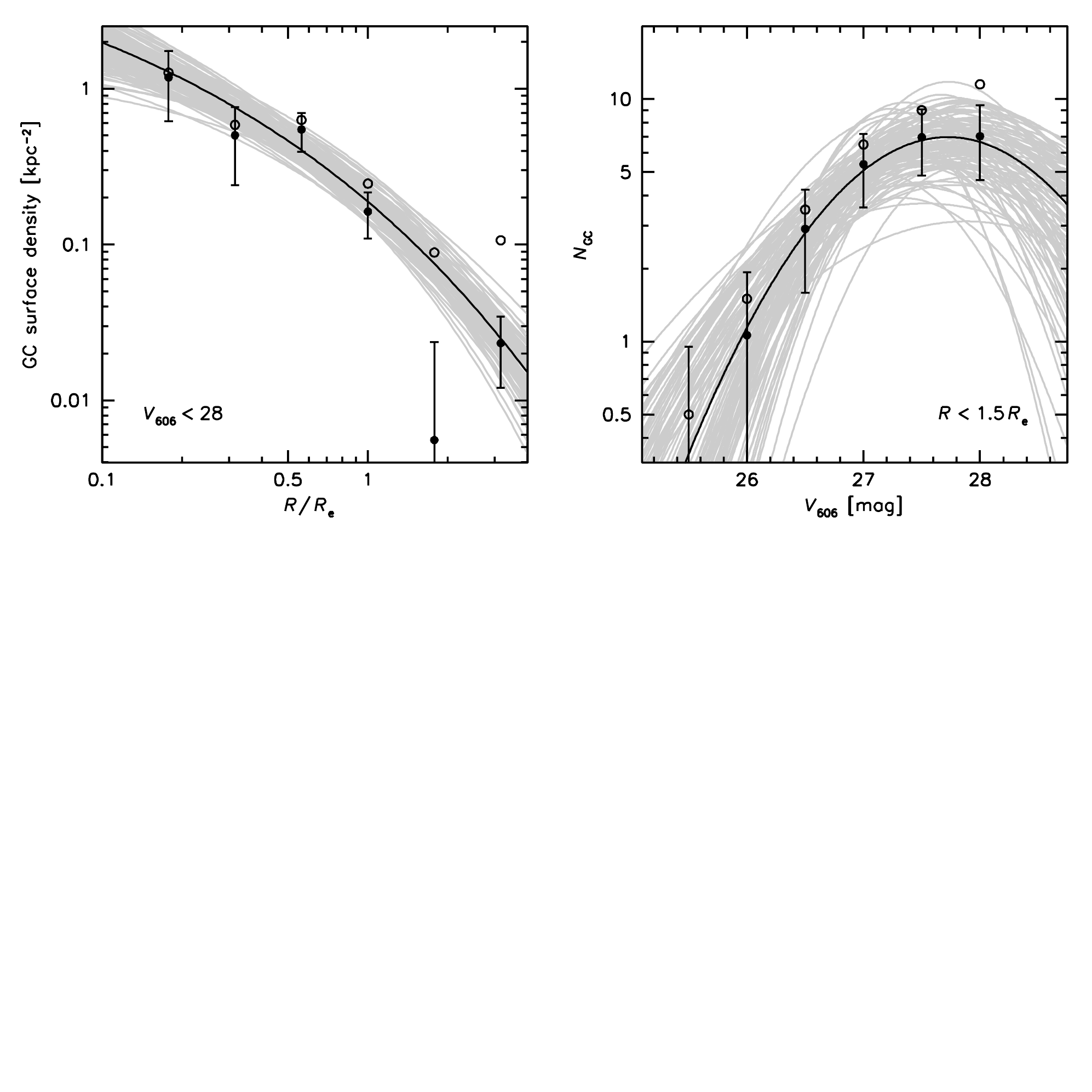}
  \end{center}
\vspace{-8.5cm}
    \caption{
{\em Left panel:} Average radial surface density profile of globular clusters
in Dragonfly\,44 and DFX1. Open circles show compact
objects with $V_{606}<28$.
Solid circles with errorbars are corrected for
contamination.
The line is the best-fit Sersic model, with
$R_{\rm gc}=2.2^{+1.3}_{-0.7}R_{\rm e}$.
Randomly drawn MCMC samples are shown in grey.
{\rm Right panel:} Average luminosity function of
globular clusters, within $R=1.5R_{\rm e}$.
The line is a Gaussian with $\langle V_{606}\rangle=
27.7^{+0.2}_{-0.2}$\,mag and $\sigma=0.82^{+0.16}_{-0.15}$\,mag.
Priors of $n<4$ and $\langle V_{606}\rangle<28$ were used in the fits.
}
\label{gc_pop.fig}
\end{figure*}

\subsection{HST Imaging}

{\em HST} imaging
for Dragonfly\,44 and DFX1 was obtained in the Cycle 24 program
GO-14643. Each galaxy was observed for
three orbits in $V_{606}$ and
one orbit in $I_{814}$, using a standard dither pattern to eliminate
hot pixels. We used ACS/WFC for DFX1 but WFC3/UVIS for Dragonfly\,44
as this enabled us to simultaneously observe Dragonfly\,42 in a parallel
ACS observation. The CTE-corrected, drizzled
images created by the STScI pipeline were used. The three $V_{606}$
images were rotated and shifted to the frame of the $I_{814}$ image and
combined. In the combination step remaining deviant pixels in the individual
images  were replaced by the average of the other two frames.
The point-source depth was measured from the rms of the
counts in empty apertures with diameter
$d=8$\,pixels, corrected to $d=\infty$
using theoretical growth curves (see {Labb{\' e}} {et~al.} 2003). We find
$5\sigma$~AB depths of $V_{606}=28.4$ and $I_{814}=26.8$ for Dragonfly\,44
and $V_{606}=27.9$ and $I_{814}=27.0$ for DFX1. The relatively modest
depth of the ACS imaging can be attributed to the now quite
severe CTE effects.

The {\em HST} images of Dragonfly\,44 and DFX1 are shown in Fig.\ \ref{image.fig};
Dragonfly\,42 is discussed in \S\,4.
The left panels are color images created from the $V_{606}$ and $I_{814}$
exposures; the right panels show the deep $V_{606}$ data at high contrast
after masking spatially-extended objects (see \S\,4).
Both galaxies are smooth and elongated, with no obvious tidal features,
spiral arms, star forming regions, or other irregularities.
%Visually,
%they resemble
%dwarf spheroidal galaxies in the Local Group such as NGC\,147, except
%that the UDGs are larger and have lower surface brightness
%(see  ).
The most striking aspect of Fig.\ 1, and the central topic of this
{\em Letter}, is the fact that
both UDGs are associated with a large number of compact
objects. For Dragonfly\,44 this was already seen in ground-based imaging,
although not as clearly (see {van Dokkum} {et~al.} 2016).
For both galaxies
the
distribution of compact objects has a broadly
similar orientation and flattening
as the smooth light. 
%Figure \ref{image.fig} confirms that at least a
%subset of UDGs have striking globular cluster populations, and are
%fundamentally different from previously-studied
%galaxies of the same luminosity.
%In the remainder of this paper we quantify this result.

\section{Globular Clusters in Dragonfly\,44 and DFX1}
\label{glob.sec}

The compact objects were identified and characterized in the following way.
First, the $V_{606}$ light of the UDGs was fit with a 2D Sersic profile,
using the GALFIT code ({Peng} {et~al.} 2002). Neighboring objects, as well as the
compact sources, were masked. The fit was done multiple times,
improving the mask in each iteration. The best-fitting Sersic model has
effective radius
$R_e = 4.7$\,kpc, central surface brightness
$\mu_{0,V}=24.1$, and Sersic index $n=0.94$ for Dragonfly\,44 and
$R_e = 3.5$\,kpc, $\mu_{0,V}=24.0$, and $n=0.90$ for DFX1. The results for
Dragonfly\,44 are in good agreement with previous ground-based measurements
({van Dokkum} {et~al.} 2015b, 2016). We also fit the $I_{814}$ data,
keeping all parameters except
the sky value and the normalization fixed to the $V_{606}$ results.
The colors of the two galaxies are the same, within
the uncertainties: $V_{606}-I_{814}=0.48 \pm 0.06$ for Dragonfly\,44 and
$V_{606}-I_{814}=0.45 \pm 0.06$ for DFX1. The total magnitudes are
$V_{606}=18.8$ and $V_{606}=19.3$ respectively.

\begin{figure*}[htbp]
  \begin{center}
  \includegraphics[width=0.8\linewidth]{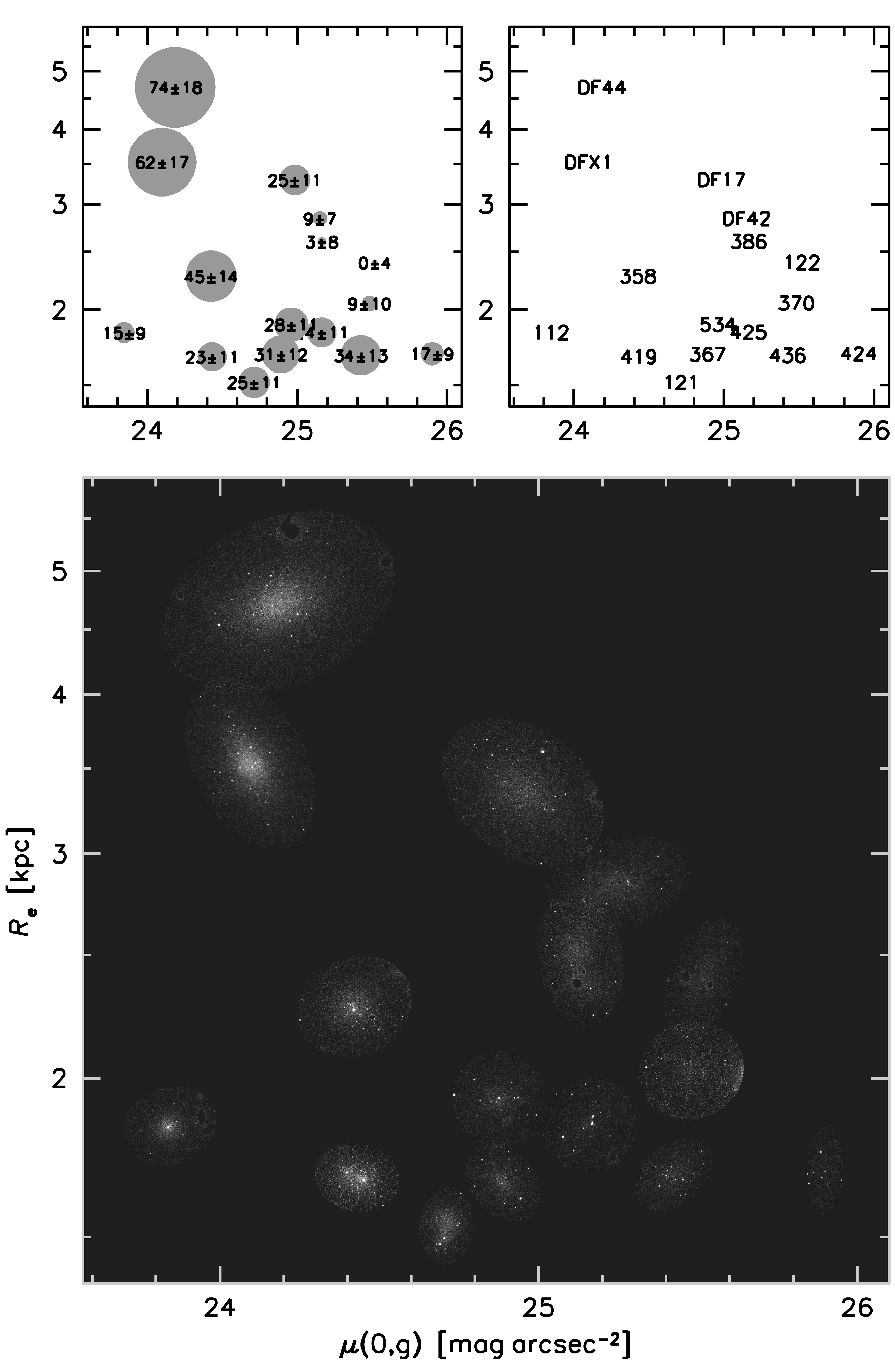}
  \end{center}
\vspace{-0.2cm}
    \caption{
Globular clusters in Coma UDGs observed with {\em HST}. The main panel
shows images of the galaxies in the plane of effective radius versus
central surface brightness, with an outer radius of $R=2R_{\rm e}$
for each cutout. Galaxies were slightly offset to minimize overlaps.
All objects in the cutouts are masked except
compact sources with $0.5<c<1.0$ and the UDGs themselves.
Many UDGs
show a larger number of compact sources than the expected
$\sim 1-3$ random ones. The top left panel shows the derived total number
of globular clusters in each galaxy, with $N_{\rm gc,tot}=4N_{\rm gc,obs}$
and $N_{\rm gc,obs}$ the contamination-corrected count within $R=1.5R_{\rm e}$
down to $V_{606}<27.6$ (see text). The size of the circles reflects
the number of clusters. The top right panel shows
galaxy identifications from van Dokkum et al.\ (2015a) or
Yagi et al.\ (2016).
}
\label{remu.fig}
\end{figure*}

After subtracting the best-fitting GALFIT models an object catalog was
created using SExtractor ({Bertin} \& {Arnouts} 1996), using default parameters.
%We verified that the
%catalog contains all visible objects in the images.
Globular
cluster candidates
were selected by the criterion $0.5<c<1.0$, where $c$ is the flux
ratio in $d=4$\,pixel and $d=8$\,pixel apertures.
Stars with a high signal-to-noise (S/N) ratio have
$c\approx 0.75$; the broad selection range ensures that unresolved objects
with low S/N are included in the sample, at the expense of some contamination
by compact galaxies.

The surface density of compact objects associated with the two galaxies
is shown in the left panel of Fig.\ \ref{gc_pop.fig}. 
The number density was measured
in elliptical annuli that are scaled to the half-light radius.
In each annulus the number of compact objects
with $V_{606}<28$ was measured and
divided by the area of the annulus (open circles). A contamination correction
was applied by subtracting the average
number density of objects with $0.5<c<1.0$,
$V<28$, and distances $R>3R_{\rm e}$.
%This
%contamination 
%is 12\,\% (Dragonfly\,44) and 16\,\% (DFX1) within $R=R_{\rm e}$.
The measurements were done separately for
Dragonfly\,44 and DFX1 and then averaged.

The radial distribution confirms the visual impression of a significant
overdensity of compact objects.  We fit a Sersic profile to the
combined, binned
distribution using the {\tt emcee} methodology ({Foreman-Mackey} {et~al.} 2013), with
a prior on the Sersic index of $n\leq 4$. The best fit has
$n=3.1^{+0.6}_{-0.9}$ and $R_{\rm gc}=2.2^{+1.3}_{-0.7}R_{\rm e}$,
with $R_{\rm gc}$ the half-number radius of the globular cluster
distribution.\footnote{We find similar results
when fitting the profiles of the
individual galaxies rather than the average.}
Forcing $n=1$, i.e., a similar functional form as the stellar
light, we find $R_{\rm gc}=1.4^{+0.2}_{-0.2} R_{\rm e}$. We conclude
that the distribution of globular clusters is more extended than the
galaxy light, as was previously found for luminous galaxies
({Kartha} {et~al.} 2014; {Hargis} \& {Rhode} 2014) and the UDG Dragonfly\,17
({Peng} \& {Lim} 2016), but that the precise value of $R_{\rm gc}$ is not
well constrained by our data.

The luminosity function of compact objects within
$R=1.5R_{\rm e}$ is shown in
the right panel of Fig.\ \ref{gc_pop.fig}. The number rises sharply
from $V_{606}\sim 26$ to $V_{606}\sim 27.5$, where it seems to plateau.
The canonical luminosity function of globular clusters is a Gaussian
with a width of $\sigma\approx 1$\,mag and a
peak at $\langle V_{606}\rangle
\approx 27.6$ for the Coma distance 
(see, e.g., {Miller} \& {Lotz} 2007; {Lee} \& {Jang} 2016; {Peng} \& {Lim} 2016). We cannot constrain
the peak magnitude very well
with our data, as the luminosity function does not show a clear turnover.
Fitting a Gaussian with a prior  $\langle V_{606}\rangle<28$, we find
$\langle V_{606}\rangle=
27.7^{+0.2}_{-0.2}$\,mag and $\sigma=0.82^{+0.16}_{-0.15}$\,mag.

We conclude that the properties of the compact objects in Dragonfly\,44
and DFX1 are consistent with
those expected from previously-studied
globular cluster populations of other galaxies.
To estimate the total number of clusters in each galaxy we
%assume that
%$R_{\rm gc}=1.5 R_{\rm e}$ and $\langle V_{606}\rangle = 27.6$. Both
%values are consistent with the observations and with previous studies,
%and enable meaningful comparisons to other galaxies.
use
%With these assumptions the number of globular clusters is simply
$N_{\rm gc} = 4 N_{\rm gc,obs}$, where $N_{\rm gc,obs}$ is the
contamination-corrected number
of compact objects with $R<1.5R_{\rm e}$ and $V_{606}<27.6$.
We find $N_{\rm gc}=74\pm 18$ for Dragonfly\,44 and $N_{\rm gc}=62\pm 17$
for DFX1. The number for Dragonfly\,44 is consistent with our previous
measurement
from ground-based imaging ($N_{\rm gc}=94^{+25}_{-20}$;
van Dokkum et al.\ 2016).

Finally, we note that the globular clusters are blue and that their
colors are similar to that of the smooth light of the UDGs, as was
previously found by
{Beasley} \& {Trujillo} (2016) for Dragonfly\,17.
Due to the limited depth of the $I_{814}$ data
we can only measure reliable colors for the brightest clusters. For
$V_{606}<26.5$ and $R<1.5R_{\rm e}$ we find $\langle V_{606}-I_{814}\rangle
=0.37\pm 0.06$.

\section{Globular Clusters in Other Coma UDGs}

%Amorisco et al.\ did not inspect or analyze the images of UDGs
%but used previously
%published catalogs to arrive at their result.
We obtained the ACS images of all
54 low surface brightness
objects from the {Yagi} {et~al.} (2016) catalog that fall in the
Coma Cluster Treasury program area from
MAST\footnote{https://archive.stsci.edu/prepds/coma/},
and analyzed these galaxies in the
same way as described above. Most of the
CCTp data consist of a single orbit
in $g_{475}$ and a single orbit in $I_{814}$. We added the images
to increase the S/N ratio, using $V'=\sqrt{2} g_{475} + I_{814}/\sqrt{2}$.
We use a zeropoint of 27.14, as for this
value derived magnitudes are
equivalent to $V_{606}$ for objects
with the colors of UDGs and their globular clusters.

Structural parameters of the galaxies were determined using GALFIT, following
the same masking procedures as described in \S\,\ref{glob.sec}. Only 12 of
the 54 
objects have $R_{\rm e}>1.5$\,kpc
and are classified as UDGs. The remaining galaxies are up to a factor of two
smaller than this limit. After subtracting the best-fitting
GALFIT models compact objects were identified, again using the same
methodology and criteria as used for Dragonfly\,44 and DFX1.
The number of globular clusters was then determined
by measuring the number of compact objects with $V'<27.6$
in an elliptical aperture with radius $R=1.5 R_{\rm e}$ and multiplying this
by 4.

\begin{deluxetable}{cccccccc}
\tablecaption{Structural Parameters and Globular Cluster Counts}
\tabletypesize{\footnotesize}
\tablewidth{0pt}
\tablehead{\colhead{Id} & \colhead{$M_V$} & \colhead{$R_{\rm e}$} &\colhead{$\mu(0,g)$} &\colhead{$n$} &
\colhead{$b/a$} & \colhead{$N_{\rm gc}$}\\
\colhead{} & \colhead{} & \colhead{(kpc)} & 
\colhead{}
 & \colhead{} & \colhead{} & \colhead{}}
\startdata
DF\,17 & $-15.3$ & 3.3 & 25.0 & 0.61 & 0.71 & $25\pm 11$ \\
DF\,42 & $-14.7$ & 2.8 & 25.2 & 0.64 & 0.61 & $ 9\pm  7$ \\
DF\,44 & $-16.2$ & 4.7 & 24.2 & 0.94 & 0.68 & $76\pm 18$ \\
DF\,X1 & $-15.8$ & 3.5 & 24.1 & 0.90 & 0.62 & $63\pm 17$ \\
Y\,112 & $-14.2$ & 1.8 & 23.8 & 1.43 & 0.81 & $15\pm  9$ \\
Y\,121 & $-14.0$ & 1.5 & 24.7 & 0.60 & 0.69 & $25\pm 11$ \\
Y\,122 & $-13.8$ & 2.4 & 25.5 & 0.64 & 0.54 & $ 0\pm  4$ \\
Y\,358 & $-14.8$ & 2.3 & 24.4 & 0.99 & 0.83 & $45\pm 14$ \\
Y\,367 & $-13.7$ & 1.7 & 24.9 & 0.84 & 0.73 & $31\pm 12$ \\
Y\,370 & $-13.9$ & 2.1 & 25.5 & 0.78 & 0.92 & $ 9\pm 10$ \\
Y\,386 & $-14.7$ & 2.6 & 25.2 & 0.53 & 0.63 & $ 3\pm  8$ \\
Y\,419\tablenotemark{a} & $-14.6$ & 1.7 & 24.4 & 0.62 & 0.78 & $23\pm 11$ \\
Y\,424\tablenotemark{b} & $-11.7$: & 1.7: & 26.8: & 0.50: & 0.41: & $17\pm  9$ \\
Y\,425 & $-13.3$ & 1.8 & 25.2 & 1.33 & 0.99 & $24\pm 11$ \\
Y\,436 & $-13.5$ & 1.7 & 25.4 & 0.58 & 0.69 & $34\pm 13$ \\
Y\,534 & $-13.9$ & 1.9 & 25.0 & 1.03 & 0.96 & $28\pm 11$ 
\enddata
\tablenotetext{a}{Y\,419 may be a superposition of two smaller
galaxies.}
\tablenotetext{b}{Y\,424 is barely detected in the HST images.}
\end{deluxetable}

\begin{figure}[htbp]
  \begin{center}
  \includegraphics[width=1.0\linewidth]{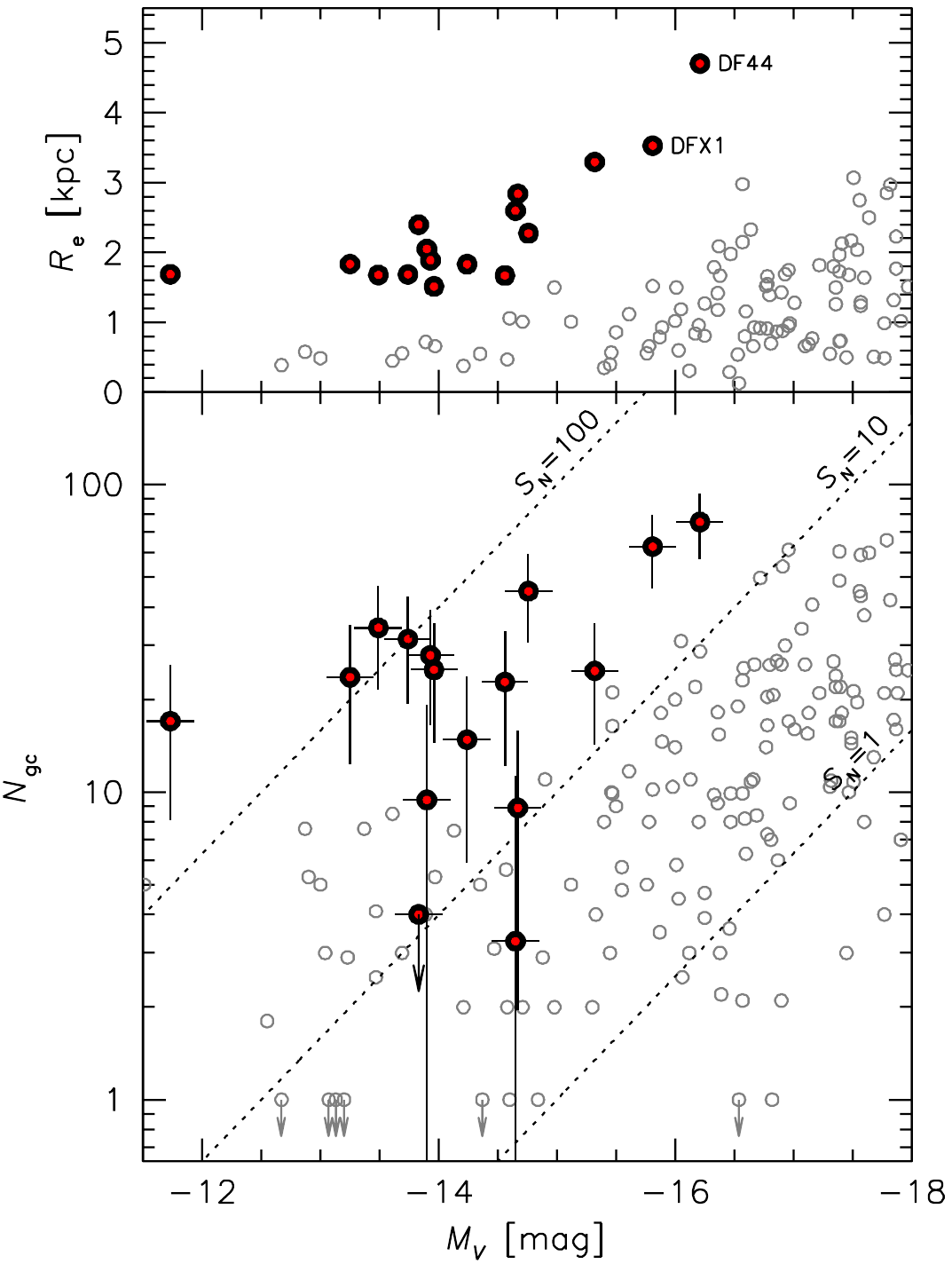}
  \end{center}
\vspace{-0.2cm}
    \caption{
{\em Main panel:}
Relation between the number of globular clusters $N_{\rm gc}$ and
total absolute magnitude $M_V$ for Coma UDGs
(solid symbols with errorbars).
Open symbols are averages for normal galaxies, derived from the
literature compilation of {Harris} {et~al.} (2013).
Broken lines indicate a constant specific frequency $S_{\rm N}$.
UDGs have $10\lesssim S_{\rm N}\lesssim 100$.
{\em Top panel:} Relation between effective radius and $M_V$.
}
\label{num.fig}
\end{figure}

\begin{figure*}[htbp]
  \begin{center}
  \includegraphics[width=0.95\linewidth]{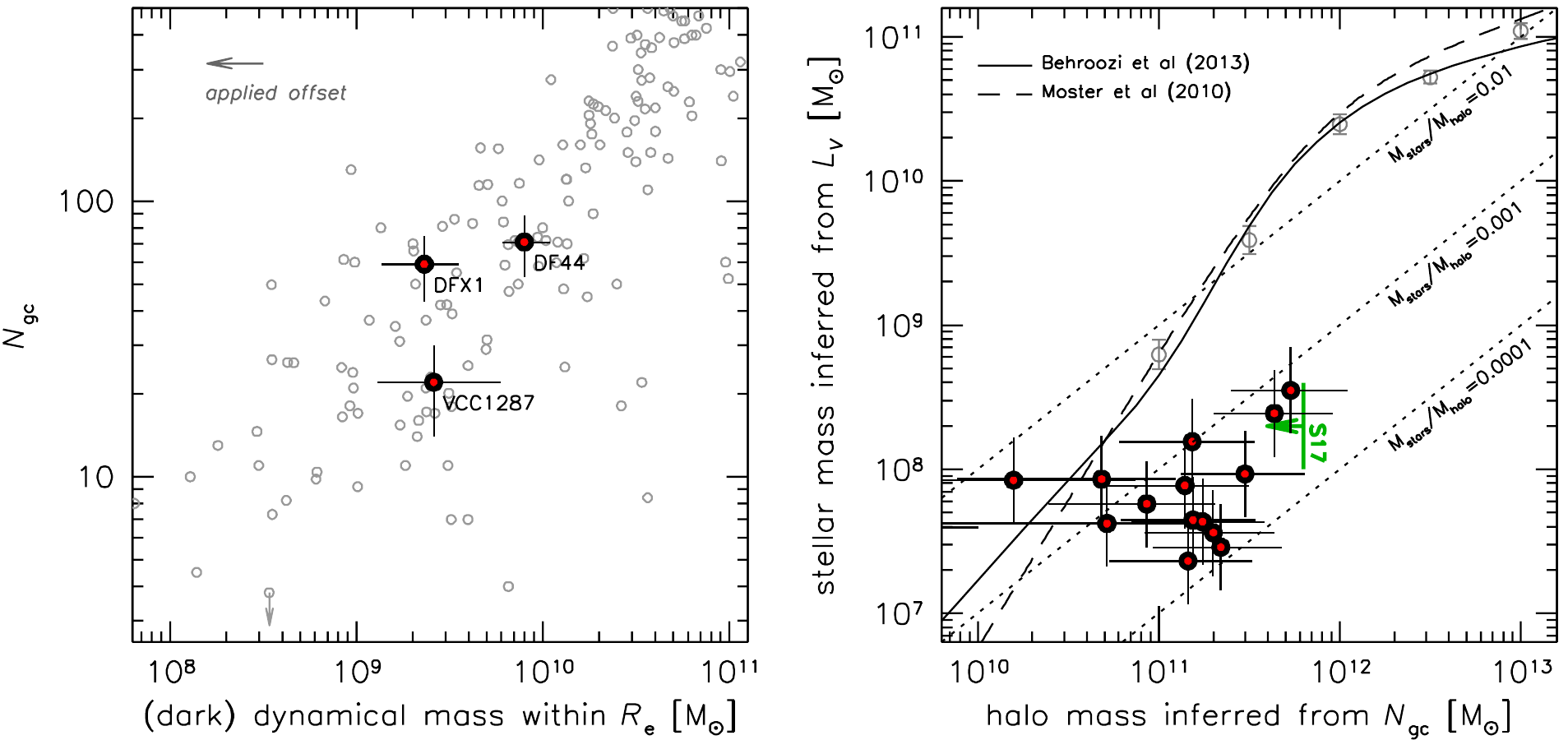}
  \end{center}
\vspace{-0.2cm}
    \caption{
{\em Left panel:}
Relation between the number of globular clusters and dynamical mass
within $R_{\rm e}$ (with $M_{\rm dyn}\propto \sigma^2R_{\rm e}$).
VCC\,1287 is
a Virgo cluster UDG from {Beasley} {et~al.} (2016). 
Grey open symbols are from {Harris} {et~al.} (2013), shifted by
0.3\,dex in mass (see text). UDGs fall on the trend defined by
other galaxies.
{\em Right panel:} Inferred stellar mass versus inferred halo mass.
The green limit labeled ``S17'' is derived from lensing ({Sif{\'o}n} {et~al.} 2017).
The solid and broken lines are derived from abundance matching
({Behroozi} {et~al.} 2013; {Moster} {et~al.} 2010).
Open symbols are normal galaxies
from {Harris} {et~al.} (2013), analyzed in the same way as the UDGs.
UDGs appear to have low stellar masses
for their halo masses, compared to previously-studied galaxies.
}
\label{halo.fig}
\end{figure*}

The results are listed in Table 1
and summarized in Fig.\ \ref{remu.fig}. Dragonfly\,44
and DFX1 have the most dramatic globular cluster populations of
all {\em HST}-observed UDGs in Coma, but they
are not the only ones with significant overdensities of point
sources.\footnote{Interestingly some
UDGs appear to be nucleated, as first reported
by {Koda} {et~al.} (2015). Here we do not attempt to distinguish
between globular clusters and compact nuclei.}
%In contrast to
%{Amorisco} {et~al.} (2017)
We find that half of the 12 {Yagi} {et~al.} (2016) UDGs have overdensities that
are significant at the $>2\sigma$ level.
Among the {Yagi} {et~al.} (2016) objects
galaxy 358 has the largest number of globular clusters, with
$N_{\rm gc}=45\pm 14$. 
We also include Dragonfly\,42, observed in parallel with
Dragonfly\,44, and Dragonfly\,17.
The globular clusters in Dragonfly\,17  were
previously studied by {Peng} \& {Lim} (2016) and
{Beasley} {et~al.} (2016). Our measurement 
is consistent with
these studies ($N_{\rm gc}=25\pm 11$ versus $N_{\rm gc}=28\pm 14$).

The globular counts in UDGs are compared to those in other galaxies
in Fig.\ \ref{num.fig}. Open symbols in this Figure are taken from
the literature compilation of Harris et al.\ (2013).\footnote{In
the luminosity regime of the UDGs the primary sources include
Miller \& Lotz (2007), Peng et al.\ (2008), and Georgiev et al.\
(2010).}
UDGs have more globular clusters than other
galaxies of the same total luminosity. For this sample of 16
UDGs the
median difference\footnote{It should be
noted that the Harris et al.\ sample is heterogeneous, and
possibly biased against galaxies with $\sim 0$ globular
clusters.} is a factor of $6.9^{+1.0}_{-2.4}$.
The specific frequency,
defined as $S_{\rm N} = N_{\rm gc} \times 10^{0.4(M_V+15)}$,
is $10\lesssim S_{\rm N}\lesssim 100$ for UDGs.

%\section{Comparison to Kinematics and Implied Halo Masses}

%scatter

%ref forbes:16, harris:17

\section{Discussion}

Using newly obtained {\em HST} images of the two UDGs in Coma with
measured kinematics we find that they have remarkable globular cluster
populations. No other known galaxies look like the objects in Fig.\ 1:
very diffuse ``blobs'' as large as the Milky Way,
sprinkled with many extremely compact sources.
Dragonfly\,44 and DFX1 are both large and relatively bright among UDGs.
Although
none of the smaller and fainter UDGs that were imaged serendipitously
in the Coma Cluster Treasury program have quite as many globular clusters,
several come close (see Fig.\ 4). Their median
globular cluster
specific frequency is actually higher than that
of Dragonfly\,44 and DFX1 ($\langle S_{\rm N}\rangle =45$ versus
$\langle S_{\rm N}\rangle =27$), because they are so faint.

Our results seem to be at odds with {Amorisco} {et~al.} (2017),
who report that UDGs in the Coma cluster do not have
a statistically-significant
excesss of compact objects compared to normal
dwarf galaxies (see \S\,1).
This tension may be partly due to the inclusion of galaxies with
$R_{\rm e}<1.5$\,kpc in that study, and partly to differences
in selection techniques: as an example,
{Amorisco} {et~al.} (2017) identify only a single compact object in galaxy 358
(N.~Amorisco, private communication),
whereas we find $13$
with $V_{606}<27.6$ and $R<1.5R_{\rm e}$ ($N_{\rm gc,obs}=11.2$ after
correcting for contamination, and $N_{\rm gc,tot}=4N_{\rm
gc,obs}=45$).\footnote{Alerted to our apparently discrepant results,
the authors of Amorisco et al.\ (2017) are revising aspects of their
analysis and it is likely that
the published version of their paper is in better agreement with
our study than the submitted version (N.~Amorisco, private
communication).}

The results presented here, and particularly Fig.\ 4,
put to rest the suggestion that most cluster
UDGs are directly related to
smaller galaxies of the same total luminosity. Although some UDGs may be
rapidly spinning low mass galaxies ({Amorisco} \& {Loeb} 2016),
and quite a few are probably tidally distorted objects on the verge
of complete disruption (see {Collins} {et~al.} 2013; {Merritt} {et~al.} 2016), the majority
appear to have a different origin.

Several studies have suggested that the number of globular clusters is more
closely related to the dark matter halo mass of a galaxy than to its stellar
content (Blakeslee et al.\ 1997;
%{Harris}, {Harris}, \& {Hudson} 2015;
{Forbes} {et~al.} 2016;
{Harris} {et~al.} 2017).
Although we cannot test this directly for UDGs, as no halo masses
out to large radius
have yet been measured, we can determine whether UDGs fall on the
same relation between $N_{\rm gc}$ and the
dynamical mass within the effective radius as other galaxies.
This relation is shown in the left panel of Fig.\ \ref{halo.fig},
with
$M_{\rm dyn}(<R_{\rm e})\approx 9.3\times 10^{5}\sigma^2R_{\rm e}\sqrt{b/a}$
({Wolf} {et~al.} 2010). Solid symbols are the
three UDGs with measured kinematics: Dragonfly\,44,
DFX1, and the Virgo galaxy VCC\,1287 ({Beasley} {et~al.} 2016).
Open symbols are from {Harris} {et~al.} (2013), after
applying an offset of 0.3\,dex  to account for the
contribution of baryons inside $R_{\rm e}$
({Grillo} 2010; {Auger} {et~al.} 2010).
\footnote{This contribution is negligible
for UDGs; see van Dokkum et al.\ (2016).}
The UDGs fall on the same relation as other galaxies.

Encouraged by this result, and following
{Peng} \& {Lim} (2016) and {Beasley} \& {Trujillo} (2016), we converted $N_{\rm gc}$
to halo mass using $\log M_{\rm halo} = 9.62 + 1.12 \log N_{\rm gc}$
({Harris} {et~al.} 2017).
The median inferred
halo mass is $M_{\rm halo}\sim 1.5 \times 10^{11}$\,\msun\ for
the sixteen galaxies; Dragonfly\,44 and DFX1 have 
inferred $M_{\rm halo}\sim
5\times 10^{11}$\,\msun.  In the right panel of Fig.\ \ref{halo.fig} we
show the relation between stellar mass
and halo mass. The stellar masses
were determined from the total magnitudes using 
{Bell} \& {de Jong} (2001), with the assumption that all UDGs have the
same $V-I$ color as Dragonfly~44 and DFX1. Open symbols are
derived from the {Harris} {et~al.} (2013) sample of normal galaxies,
using  their $V-K$ colors 
to transform luminosity to mass.
The UDGs fall below the canonical relations between stellar
mass and halo mass,
suggesting they are ``failed'' galaxies that quenched after
forming their globular clusters but before forming a disk and bulge
(see also {Peng} \& {Lim} 2016; {Beasley} \& {Trujillo} 2016; {van Dokkum} {et~al.} 2016).

This study can be extended and improved in various ways.
More
dynamical measurements are needed to test whether the globular cluster
counts are indeed directly related to the dark matter content, and to
test whether there is a simple relation between the structure
and kinematics of UDGs ({Zaritsky} 2017).
The UDGs that overlap with the CCTp program are relatively small
-- none were in the original Dragonfly sample -- and {\em HST}
imaging of more UDGs with $R_{\rm e}\gtrsim 3$\,kpc may turn up
even more spectacular objects than Dragonfly\,44 and DFX1.
%The
%Cycle 25 program GO-14658 (PI: E.~Peng) should address this question.
The best way to measure total masses is probably through
weak lensing. A recent ground-based study has provided the first
upper limits ({Sif{\'o}n} {et~al.} 2017, see Fig.\ 5),
and future {\em HST} studies of large samples
could probe deeper into the relevant
halo mass range of $10^{11}$\,\msun\ -- $10^{12}$\,\msun.

\acknowledgements{We thank the anonymous
referee for a constructive report
and for independently inspecting the HST images.
Support from {\em HST} grant GO-14643 and
NSF grants AST-1616598, AST-1616710, and
AST-13123761 is gratefully acknowledged.
AJR is a Research Corporation for Science Advancement Cottrell Scholar.}

%% --------------------------------------------------------------------
%% Tue May 23 13:09:00 2017
%%   This file was generated automagically from the files
%%   coma_udgs.bbl and coma_udgs.tex using
%%     nat2jour.pl
%%   This file should accompany coma_udgs-aas.tex.
%% --------------------------------------------------------------------

\end{document}